# Nonlinear interferometry with infrared metasurfaces

Anna V. Paterova*, Dmitry A. Kalashnikov*[+], Egor Khaidarov, Hongzhi Yang, Tobias W. W. Mass, Ramón Paniagua-Domínguez, Arseniy I. Kuznetsov, and Leonid A. Krivitsky[++]

*Institute of Materials Research and Engineering, A\*STAR (Agency for Science Technology and Research), 138634 Singapore*

\* *equal contributions*

[+]Dmitry_Kalashnikov@imre.a-star.edu.sg

[++]Leonid_Krivitskiy@imre.a-star.edu.sg

The optical elements comprised of sub-diffractive light scatterers, or metasurfaces, hold a promise to reduce the footprint and unfold new functionalities of optical devices. A particular interest is focused on metasurfaces for manipulation of phase and amplitude of light beams. Characterisation of metasurfaces can be performed using interferometry, which, however, may be cumbersome, specifically in the infrared (IR) range. Here, we realise a new method for characterising IR metasurfaces based on nonlinear interference, which uses accessible components for visible light. Correlated IR and visible photons are launched into a nonlinear interferometer so that the phase profile, imposed by the metasurface on the IR photons, modifies the interference at the visible photon wavelength. Furthermore, we show that this concept can be used for broadband manipulation of the intensity profile of a visible beam using a single IR metasurface. Our method unfolds the potential of quantum interferometry for the characterization of advanced optical elements.

**Introduction.**

Planar optical elements based on sub-wavelength light scatterers, referred to as metasurfaces, have gained significant interest over the last few years. They can introduce abrupt local changes in the amplitude, polarization, and phase of a light wave, either actively or passively [1-7]. Stemming from the sub-wavelength size of their constituting parts, they are capable of modifying wavefront profiles with much higher spatial resolution and smaller form factor than conventional diffractive optical elements and spatial light modulators (SLMs) [7-10]. In particular, metasurfaces made of high refractive index dielectric materials benefit from low optical loss, compatibility with industrial fabrication processes (e.g. CMOS), and distinctive resonance behaviour [11, 12]. That is why metasurfaces hold a promise to serve as enabling components for the next generation of optical devices for augmented and virtual reality, microscopy, imaging, optical communications, and many others.

One of the main use of metasurfaces is the control of the phase profile of a beam. Characterisation of the phase encoded by such metasurfaces can be performed e.g. by interferometry [8, 13, 14]. In these experiments, the probe beam, which reflects from (or passes through) the metasurface, is overlapped with a reference beam. The resulting interference pattern reveals the phase distribution imposed by the metasurface onto the incident beam. While interferometric measurements in the visible range are rather straightforward, measurements in the infrared (IR) have practical challenges. They are associated with the limited efficiency and tunability, high noise, and high cost of IR light sources and array photodetectors. With the rapid developments and deployment of IR metasurfaces, there is an urgent need for new methods for their accurate characterization.

There is a growing interest in the use of methods of nonlinear and quantum optics for studying structured beams. An interesting approach is based on nonlinear frequency mixing using the second harmonic generation (SHG) or the sum-frequency generation (SFG) processes. Both methods bring detection from IR to visible range [15-18]. These techniques involve two beams from two IR lasers (SFG) or a split beam of the IR pump (SHG), where one beam is structured, while another one is used as a reference. The two beams are then combined at the nonlinear crystal so that the resulting SFG or SHG signal carries the information about the phase or amplitude profile of the structured beam. While the method negates the need for the IR array detector, it still requires relatively powerful pulsed IR-range laser. Furthermore, measurements at different wavelengths require sophisticated tunable lasers. The use of metasurfaces in quantum optics has also been demonstrated in several recent works [19-23]. They include novel schemes for efficient generation, manipulation, and measurement of entangled states of light.

Here, we develop a new approach for characterisation of metasurfaces, and nonlocal manipulation of the angular momentum of single photons, based on the nonlinear interference of correlated photons, also referred to as *induced coherence* [24-25]. This method allows assessing the sample properties in the challenging for detection and broadband IR range by using an accessible light source and a photodetector for visible light. Earlier, nonlinear interferometers have been used for several metrological applications, including IR imaging, spectroscopy, optical coherence tomography, and polarimetry [26-35]. Here we apply it for the characterisation of IR dielectric metasurfaces and revealing fabrication imperfections. We also demonstrate that metasurfaces designed for the IR range can bring their functionality into visible by generating intensity patterns at a few-photon level in the visible beam.

We generate correlated photon pairs in a nonlinear crystal with one photon in the visible (signal), and the correlated one in the IR range (idler), via Spontaneous Parametric Down Conversion (SPDC) [24, 25]. The photons are then launched into a balanced Michelson interferometer, where they are split by a dichroic beamsplitter. The IR photons interact with the metasurface under test, while visible and pump photons are reflected by a plane mirror. All the photons are then sent back to the nonlinear crystal, where the pump generates photon pairs, identical and coherent with those launched into the interferometer. The state-vectors of the photon pairs created in the first and the second passes of the pump through the crystal interfere. The interference leads to the modulation of the intensity of the detected visible photons. The phase governing the modulation depends on the phases of all the three participating photons: visible, IR, and pump. Thus, from the observation of the interference pattern of the visible photons, we can infer the information about the phase induced by the metasurface on the IR photons. We can then assess the quality of the fabricated metasurface for verification of the design and tracing the fabrication accuracy.

The same setup can be used for characterisation of metasurfaces at different wavelengths with a change of the phase-matching conditions in the nonlinear crystal, for example by changing the crystal temperature. This is a practical and economic advantage of our experiment with respect to the alternative nonlinear methods [15-18], which would require the use of expensive tunable lasers.

Another exciting feature of our experiment is that a single metasurface in the IR range can 'virtually' shape the visible beam at the few-photon level at multiple wavelengths. This is achieved by tuning the wavelength of the pump beam, while the wavelength of the probing IR photons is fixed to match the range of the optimal performance of the metasurface. In general, the tunability of the method is limited only by the transparency and phase matching conditions of the nonlinear crystal. Thus our method opens up the possibility to expand the range of operations of metasurfaces designed for IR into the visible spectral range avoiding undesired absorption losses.

To prove our concept, we characterize metasurfaces designed to produce vortex and Laguerre Gaussian IR beams. We then show that the method allows generating annular and doughnut-shaped beams in the visible range using silicon metasurfaces operating at the near-IR wavelengths [36]. We also demonstrate the intensity modulation of the visible beam at different wavelengths using a single metasurface designed for IR.

## Results

**Metasurface design**

Our metasurfaces are made of silicon (Si) nanocylinders of 650 nm height, which are fabricated on top of an optically thick aluminum layer, acting as a mirror, on a $SiO_2$ substrate. The nanostructures are separated from the mirror by a 500 nm $SiO_2$ dielectric spacer (see Fig. 1a). The phase retardation induced by Si cylinders of different diameters (from 200 nm to 400 nm) at the operational wavelength of 1550 nm is calculated assuming that they form regular array with the period of 650 nm (see Methods for details on the numerical simulations). This range of diameters provides the required 0-$2\pi$ range in the phase retardation to allow mapping of any desired wavefront while keeping reflectivity values above 85% (see Fig. 1b).

Based on these results, we design and fabricate metasurfaces to generate complex beam profiles with desired azimuthal and radial variations. To do so, we map the phase distribution of the incident beam, which we assume to be a plane wave impinging normally to the metasurface, to the phase profile of the desired beam. We start by generating vortex beams, for which, ignoring the polarization, the transverse field distribution can be described as $E(r,\varphi) = F_m(r)exp[il\varphi]$. There are two factors in this expression. The first factor, $exp[il\varphi]$, defines the azimuthal variation of the beam, characterized by the azimuthal index $l$ (sometimes referred to as the topological charge of the angular momentum of the beam) and the azimuthal angle $\varphi$. The second factor, $F_m(r)$, accounts for the radial variation, and it is determined by the radial index $m$. Besides the vortex beam, we also apply our technique to generate a Laguerre Gaussian beam and, finally, a different class of beams, which we refer to as *annular beams*, where the intensity distribution within the rings is given as $I(r,\varphi)$, with $r$ and $\varphi$ being the polar coordinates. In these beams, the azimuthal order is zero, and the radial part is different from that of usual vortex beams [36]. The phase profiles of all these beams are shown in Fig. 2 (a, c, e, g). To fabricate the metasurfaces, we use Electron Beam Lithography (EBL) followed by Reactive Ion Etching (RIE), as detailed in the Methods section. The corresponding SEM images of the fabricated samples are shown in Fig. 2 (b, d, f, h).

**Nonlinear interferometer**

We use nonlinear interferometry to, first, characterize the response of IR metasurfaces using visible light and, second, to generate complex beams in the broad visible range using the same IR metasurface. The scheme of the nonlinear interferometer setup is shown in Fig. 3 (see Methods for the detailed schematics and description). The frequency-nondegenerate SPDC occurs in the nonlinear crystal, where the phase-matching condition is chosen in such a way that the wavelengths of signal (detected) and idler (probe) photons are in the visible and IR range, respectively [28-33]. The photons are sent into the interferometer, where they are separated by a dichroic mirror DM. Signal and pump photons are reflected by the reference mirror M and the idler photon is reflected by the metasurface under study. The confocal three-lens system in each arm of the interferometer projects photons on the metasurface and the reference mirror [33]. The reflected pump passes through the crystal for the second time and, with some probability, generates another pair of photons, which is identical and coherent with those launched initially into the interferometer.

When one cannot *in-principle* distinguish if the photon pairs were generated in the first or in the second pass of the pump through the nonlinear crystal, the interference is observed. This interference is associated with the effect of induced coherence without induced emission, discovered by Zou, Wang, and Mandel [24, 25]. We emphasize here that this effect is associated with the interference of probability amplitudes (wavefunctions) of down-converted pairs, rather than with the interference of real fields. The intensity dependence observed at the signal photon wavelength is given by [30, 31]:

$$I_s \propto \left(1 + |r_i||\tau_i^2||\mu(\Delta t)|\cos(\varphi_i + \varphi_s - \varphi_p)\right) \tag{1a}$$

$$|\mu(\Delta t)| = 2\pi \int_0^\infty d\Omega |S(\Omega)| e^{-i\Omega \Delta t}, |\mu(0)| = 1, \tag{1b}$$

where $\tau_i$ is the amplitude transmission coefficient of idler photons in the interferometer; $r_i$ is the amplitude reflection coefficient of idler photons by sample surfaces; $|\mu(\Delta t)|$ is the normalized first-order correlation function of the SPDC field; $\varphi_{i,s,p}$ are the phases acquired by the signal, idler and pump photons respectively; $\Delta t$ is the time delay between signal and idler photons in the interferometer; $|S(\Omega)|$ is the spectrum of SPDC photons; $\Omega$ is the frequency detuning. From Equation (1a) it follows that losses and phase changes, experienced by idler photons due to interaction with the metasurface, are revealed in the interference pattern of the signal photons. Hence, the observed nonlinear interference allows retrieving information about the metasurface properties at the idler photon wavelength (IR in this case) by measuring the interference pattern for signal photons only (visible in this case), i.e. direct detection of idler photons is not required. While, in this particular case, we use this technique to study the spatial distribution of the phase imparted by the metasurface, one can envision scenarios in which modulation of the polarization or amplitude are studied using the same concept.

**IR metasurfaces characterized by visible light**

We first calibrate our setup by substituting the metasurface with the mirror. The measured visibility of the interference pattern constitutes V=63±1.4% (see *Supplementary Figure S1*). Next, we introduce the metasurfaces in our interferometer. Altogether we have characterized four different metasurfaces numbered as follows: 1) the annular beam structure (Fig. 2 a, b), 2) the vortex structure with topological charge $l$=2, m=2 (Fig. 2 c, d), 3) the Laguerre-Gauss structure with $l$=2, m=1 (Fig. 2 e, f), and 4) the vortex structure with topological charge $l$=6, m=1 (Fig. 2 g, h).

We perform a fine scan of the phase in the interferometer (see Methods) and observe the modification of the interference pattern, as the light illuminating the sample acquires different phases, see the animation.

First, we analyze the intensity distribution for the metasurfaces producing vortex and Laguerre-Gauss beams. The summary of our results is shown at Fig. 4. Figures 4 a, c, e show theoretical calculations of the intensity distributions of interference patterns in the IR range. The calculated profiles are obtained considering the interference of the IR beam, modified by the metasurface, with the Gaussian beam. The figures 4 b, d, f show the experimental interference patterns measured in the visible. As we can see from Fig. 4, the patterns contain several radial beams, which correspond to the topological charge induced by the metasurface, while the degree of their chirality corresponds to the radial index. For the case of metasurfaces fabricated to produce vortex beam (structures 2 and 4), we infer that the topological charges $l$ induced by the metasurface are $l$=2 and $l$=6, and radial indexes $m$=2 and $m$=1, respectively. We also obtain topological charge $l$=2 and radial index $m$=1 for the Laguerre-Gauss structure (structure 3).

Our results reveal several peculiar features of the fabricated metasurfaces. First is the phase step-like gradients (see Fig. 4d), which are related to discretization and the rounding up (~5 nm) of the phase to cylinder diameter mapping in the lithography process. Second, the phase nonuniformity at the edge of the structure in Fig. 4f is due to different levels of EBL exposure doses required to produce cylinders of the same size at the edge and the center of the metasurface. Thus we show that our method can indeed reveal the fabrication irregularities in the studied metasurfaces.

Next, we analyze the intensity distribution in the case of annular beam shaping metasurface, see Fig. 5 a, b. We analyse two cases: when the phase acquired in the interferometer is equal to $2\pi n$ and when it is equal to $\pi+2\pi n$. We take the cross-cut from the experimental intensity plot, measured at 810 nm (red), and compare it with the theoretically calculated intensity distribution of the metasurface at 1550 nm (black), see Fig 5 e, f. For all the cases, the nearly perfect agreement between our experiments and the theory shows that (1) our technique can indeed be used for characterizing IR metasurfaces using light sources and detectors for the visible range and (2) the fabricated metasurfaces are of acceptable quality.

**Shaping the visible light beams at different wavelengths by a single metasurface**

Another exciting feature of our experiment is that a single metasurface designed for the specific IR wavelength can be used to shape visible beams (at a few photon level) at multiple wavelengths. In our experiment, described above, we design a metasurface for a specific IR wavelength and show that it is possible to use it to shape a visible light beam at 810 nm. Here, by simply tuning the pump wavelength and adjusting the phase-matching conditions in the crystal, we keep the wavelength of idler photons at the operation wavelength of the metasurface while tuning the wavelength of the visible beam. Thus, the same metasurface "virtually" shapes the visible light at multiple wavelengths, which is limited by the transparency range of nonlinear crystal and requirements of phase-matching conditions.

We demonstrate this idea with the metasurface designed for annular beams. We change the detected wavelength of the visible photon from 810 nm to 760 nm by shifting the pump wavelength from 532 nm to 514 nm and adjusting the phase-matching conditions by changing the temperature of the nonlinear crystal. We measure the intensity distribution for phase differences $2\pi n$ and $\pi+2\pi n$ in the interferometer, see Fig. 5 c, d. As in the previous case, we compare the cross-cuts of the experimental data (green) with theoretically calculated distributions (black), see Fig. 5 e, f. Based on the comparison presented in Fig. 5 e, f the results measured at 810 nm (red) and at the tuned wavelength of 760 nm (green) are found to be in good agreement with each other, and with theoretically calculated intensity distributions for 1550 nm. From here, we can conclude that using our method metasurfaces designed for IR can indeed be used for beam shaping at multiple wavelengths in visible. Interestingly, the intensity distribution in the visible beam can also be manipulated by shifting either the metasurface or the mirror for signal or pump photons along the interferometer arms, allowing, for example, to transform a high-intensity region to a low-intensity one, see panels a - d in Fig. 5.

## Discussion.

We have demonstrated a technique for characterising metasurfaces designed to operate in the IR range using only visible range sources and detectors. The concept is based on the nonlinear interference of correlated photons. It can be extended to other frequency ranges by simple adjustment of the nonlinear crystal, thus negating the need for tunable lasers. We show that our method reveals irregularities in the fabricated metasurfaces, which can find direct applications for quality assurance (QA) and quality control (QC) of the fabrication processes.

Furthermore, we have shown that the same approach can be used to modulate the intensity of a visible beam across multiple wavelengths using a single metasurface designed to operate in the IR. Though we have shifted the wavelength of the visible photon, the idler photon still stays within the optimal operation range of the metasurface. This method can be used for beam shaping at different wavelengths in the visible range using a single metasurface. It may open exciting opportunities for dynamic optical manipulation in time-varying optical traps and low-light microscopy using low-loss and easy to fabricate IR-optimized metasurfaces

Our method can be extended beyond the near-IR spectral ranges. With the appropriate choice of the nonlinear crystal, it can be further extended up to mid-, far-IR and THz ranges [37, 38]. Furthermore, a similar concept can be realised in the high-gain parametric down-conversion, allowing the manipulation of the intensity of relatively bright beams [39]. It also has practical appeal for QC and QA in the scalable manufacturing of flat optics. We believe that the method might be useful for the development of novel quantum-inspired imaging techniques, bringing together capabilities in quantum optics and nanophotonics for developing new devices and methods.

## Methods

**Numerical simulations**

To compute the reflectivity and the phase accumulation provided by the metasurface elements, we simulated a single unit cell using periodic boundary conditions along with the transverse directions and normally-incident plane wave excitation. Perfectly matched layers (PMLs) are used above the Si cylinders and in the $SiO_2$ layer (thus considered semi-infinite in our simulations). The diameter of cylinders was varied from 200 to 400 nm while keeping a constant period of 650 nm. Amorphous silicon material parameters used in the simulations correspond to those measured by ellipsometry on our deposited films, closely following those reported in the literature [40]. All simulations were performed using a Finite Difference Time Domain-based commercial solver (Lumerical FDTD), targeting an optimal performance at 1550 nm wavelength.

**Experimental realization**

The detailed schematics of the nonlinear interferometry setup is presented in Fig. 6. In the experiment, we use the tunable continuous wave (CW) laser (C-WAVE Hubner Photonics) (see Fig.6). The pump at 532 nm produces SPDC photons at 810 nm (signal) and 1550 nm (idler) wavelengths in the periodically polled lithium niobate (PPLN) crystal with the length of 10 mm and with 7.5 µm poling period, heated at 70 ˚C. To produce the signal photons at 760 nm and keep the idler at 1550 nm, we tune the pump wavelength to 514 nm and adjust the phase-matching conditions by changing the poling period to 6.81 µm and setting the temperature of PPLN crystal at 79˚C. The generated SPDC photons are separated into different arms of the interferometer by a dichroic mirror $D_2$ (Semrock): visible and pump photons travel in one arm, and IR photons travel in another arm. In each arm of the interferometer, we insert the three-lens system to achieve the required spatial resolution of the phase mapping. The system consists of three BK7 lenses, where the first two lenses $F_1$ and $F_2$ have $f_{1,2} = 75$ mm focal lengths, and the lens $F_3$ has $f_3 = 5$ mm. The system provides a spatial resolution of 12.4 µm (see *Supplementary Figure S2*). Then, all the beams are reflected in the crystal: visible beams by the reference mirror and the IR beam by the metasurface. The reflected pump beam generates another pair of SPDC photons. Their interference pattern in the visible range is observed by a standard CMOS camera (Thorlabs CS2100M-USB) with a pixel size of 5.04 µm.

The sample is mounted onto a motorized XYZ translation stage (Thorlabs) and unidirectional piezo stage. The XY translation allows us to image different areas of the sample. The coarse Z translation allows for balancing interferometer arms. The piezo stage provides fine-tuning along the Z direction. The optimal position is found at the point which corresponds to the highest interference visibility, defined as $V = \frac{I_{max}-I_{min}}{I_{max}+I_{min}} = \frac{A}{y_0}$, where $A$ is the amplitude of modulation for interference pattern, and $y_0$ is the base signal without interference. Then, the sample is scanned in a step of ~20 nm with the piezo stage. Each position of the piezo stage is related to a phase image of the structure. The typical acquisition time for one image is 300 ms.

To analyze the data, we offset the reference visibility value obtained with the mirror and infer the intensity distribution introduced by the IR metasurface to the visible beam.

**Acknowledgments**

We acknowledge the support of the Quantum Technology for Engineering (QTE) program of A*STAR project № A1685b0005 and the A*STAR SERC Pharos programme (grant number 152 73 00025, Singapore).

**Author contributions**

DAK, AVP, RPD, and LAK jointly conceived the idea of the experiment. AVP and HY build the nonlinear interferometer and conducted optical measurements. EK, TWM, RPD, AIK designed, simulated, and fabricated the metasurfaces. AVP and DAK analysed the experimental data. DAK wrote the first draft of the manuscript with the contributions of all co-authors. DAK and LAK coordinated the project.

**Competing interests**

The authors declare no competing financial interests.

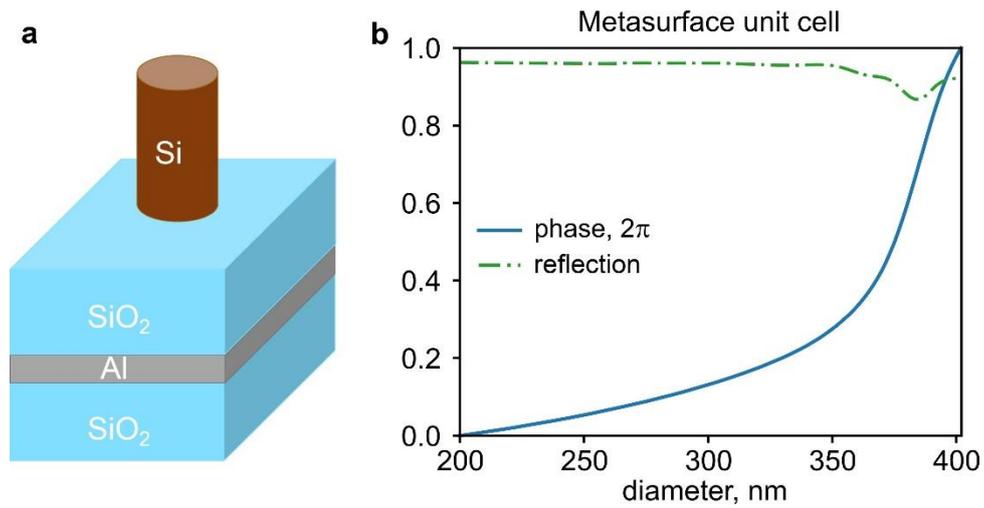

**Figure 1**. **a)** A unit cell of metasurface consisting of Si cylinder and $SiO_2$ + Al + semi-infinite $SiO_2$ substrate. **b)** The dependence of the acquired phase and reflection of a regular array of Si cylinders with 650 nm period on the cylinder diameter

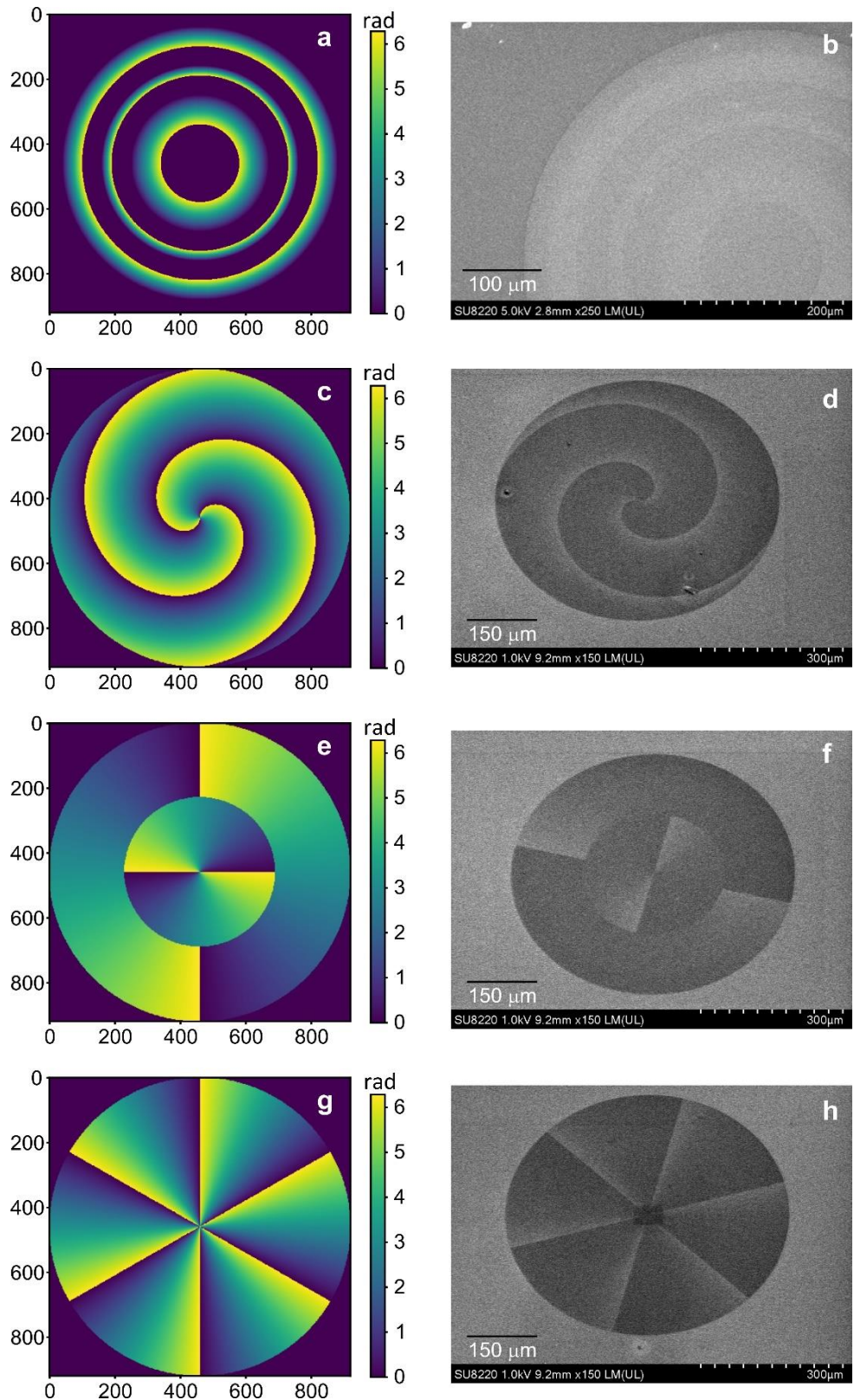

**Figure 2.** Theoretically calculated phase distributions and SEM images for metasurfaces producing: **a** and **b**) annular beam, **c** and **d**) vortex with topological charge *l*=2, m=2, **e** and **f**) Laguerre-Gaussian beam with *l*=2, m=1 **g** and **h**) vortex with topological charge *l*=6, m=1.

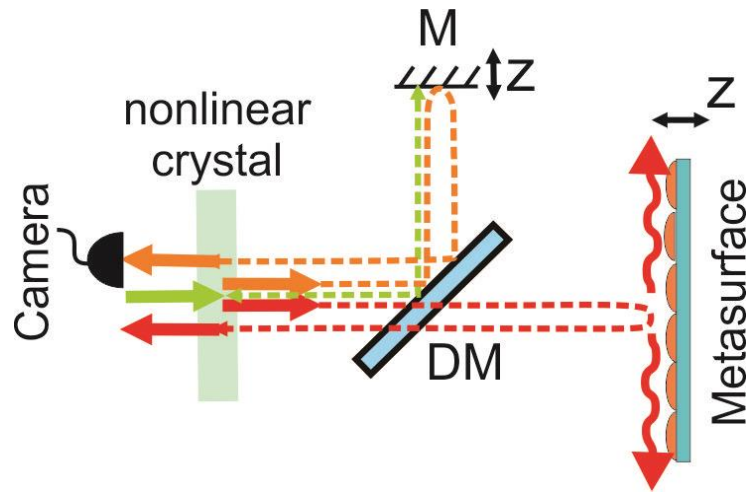

**Figure 3.** The schematic of the nonlinear interferometer. Arrows indicate the interfering photons; dashed lines show the paths of the photons. The laser (green arrow) generates a signal (visible, orange arrow) and idler (IR, red arrow) photons in the nonlinear crystal. Photons are split by a dichroic mirror (DM). Pump and signal photons are reflected by a mirror (M); idler photons are reflected by a sample under study. The interference of visible photons is detected by a camera as a function of displacement Z of mirror M or metasurface itself. Properties of the metasurface in the IR range are inferred from the interference of photons in the visible range. Green, orange and red arrows stand for the pump, signal and idler photons, respectively. The beams are shifted for clarity.

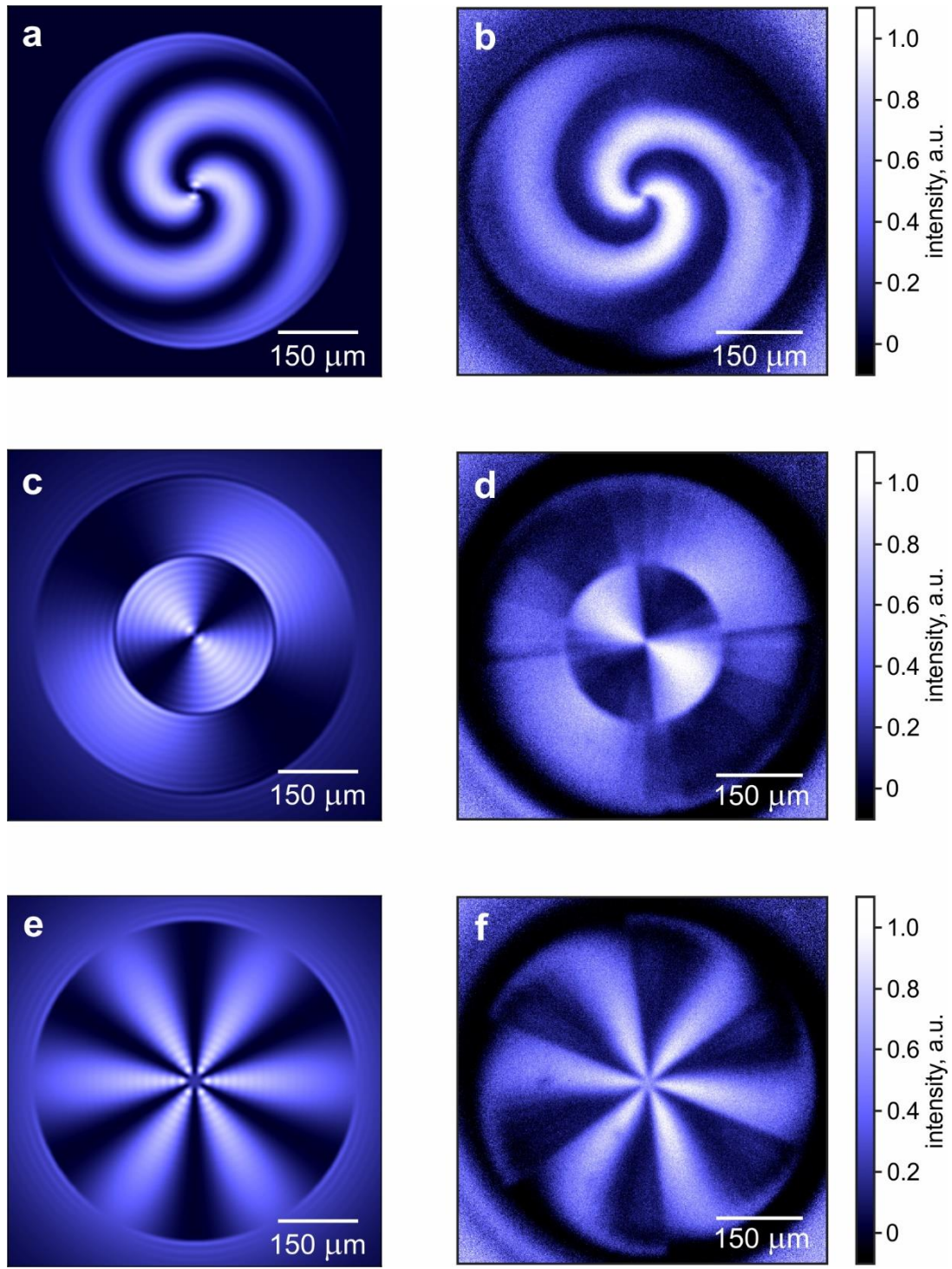

**Figure 4.** Calculated at 1550 nm and measured at 810 nm interference patterns for **a** and **b**) vortex structure with topological charge *l*=2, m=2, **c** and **d**) Laguerre-Gauss structure with *l*=2, m=1, **e** and **f**) vortex beam structure with topological charge *l*=6, m=1.

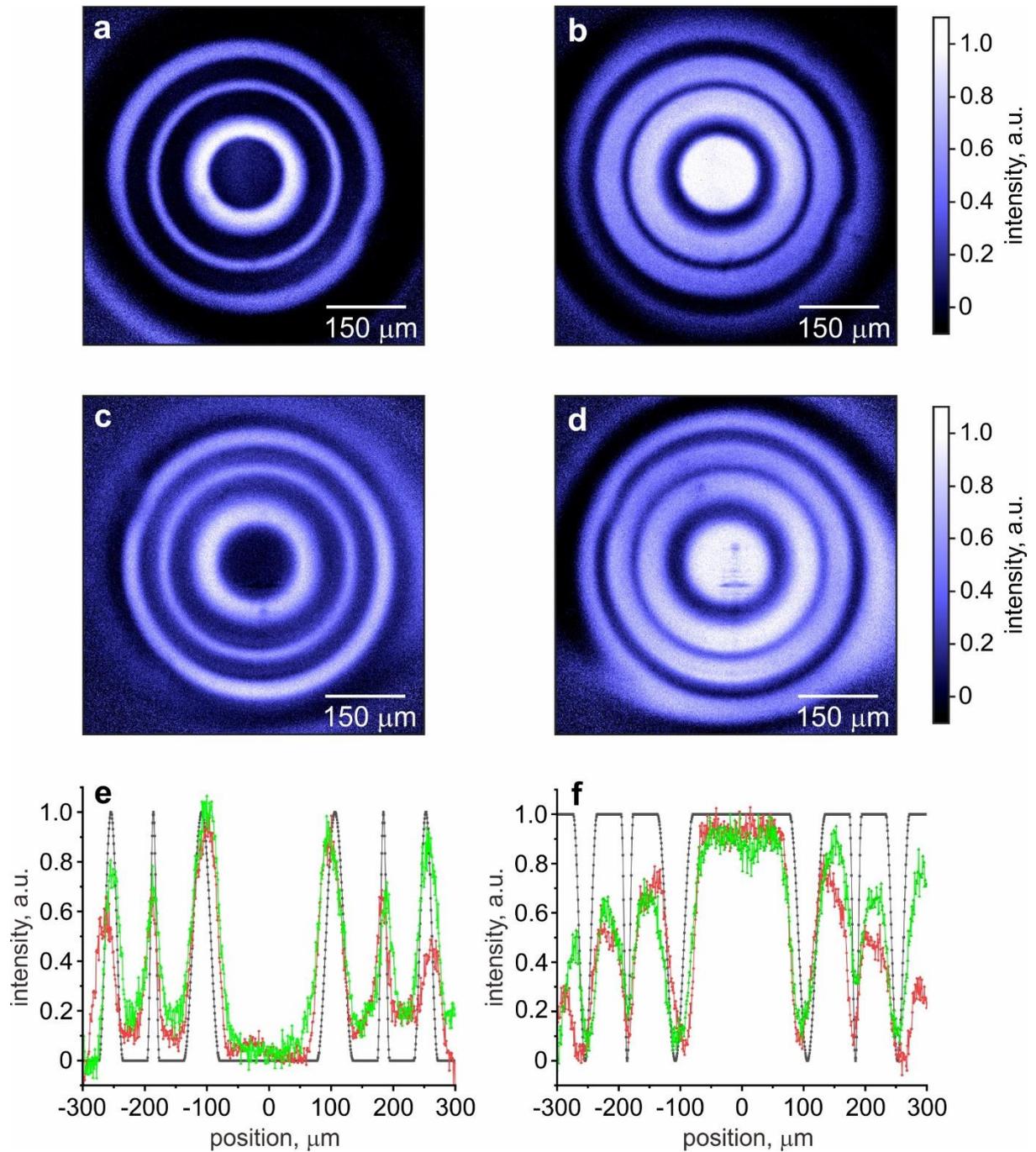

**Figure 5.** Intensity distribution for annular beam structure. **a** and **b)** Interferograms obtained at 810 nm when the phase in interferometer is equal to **a)** $\pi+2\pi n$ and **b)** $2\pi n$, **c** and **d)** Interferograms obtained at 760 nm when the phase in interferometer is equal to **c)** $\pi+2\pi n$ and **d)** $2\pi n$, **e** and **f)** Cross-cuts of the obtained interferograms at 810 nm (red) and 760 nm (green) compared with theoretical intensity distribution at 1550 nm (black).

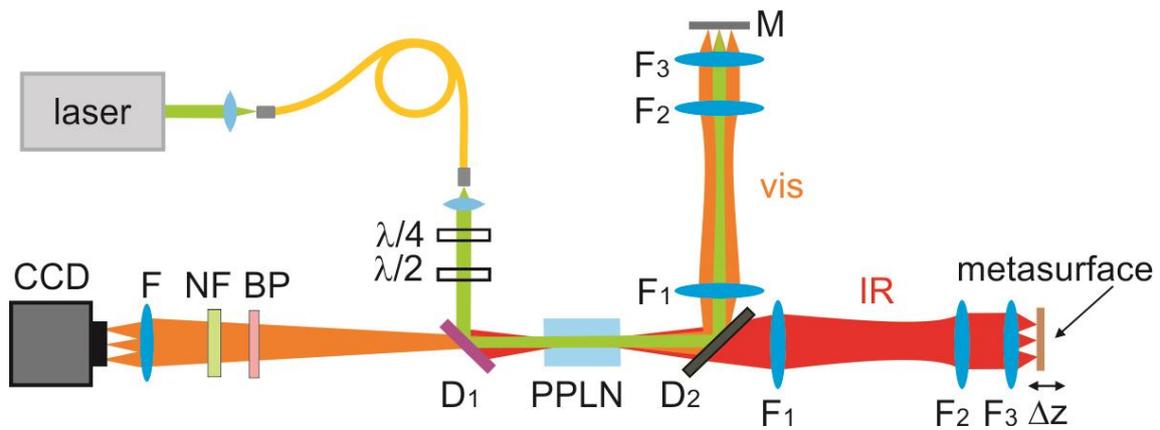

**Figure 6.** Experimental Setup. The tunable continuous-wave (cw) laser, injected through a dichroic mirror D1, is used as a pump source for the PPLN crystal. The visible and IR photons are separated into different channels by a dichroic beamsplitter $D_2$. The visible and pump photons are reflected by the mirror M, while the IR photons are reflected by the metasurface. The metasurface is mounted on the motorized XYZ translator for the coarse and the piezo stage for the fine movement of the sample. The interference of signal photons is detected by a standard visible light CMOS camera (Thorlabs). The signal is filtered by the notch (NF) and bandpass (BP) filters. XY axis corresponds to the surface plane of the sample.

**Supplementary Materials.**

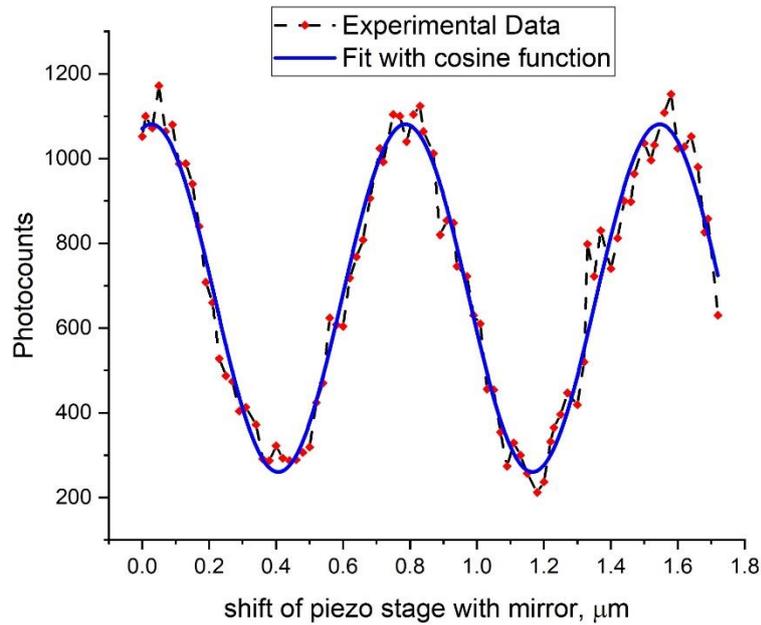

**Figure S1.** Reference visibility of interference pattern in visible measured without a sample fitted with cosine function in Eq. 1(a), giving $A=420\pm9$ and $y_0=670\pm6$ with $COD(R^2)\approx0.97$. The data presented gives the visibility value of $V=63\pm1.4\%$ for the measured reference visibility, which was subsequently offsetted at the measurements with metasurfaces.

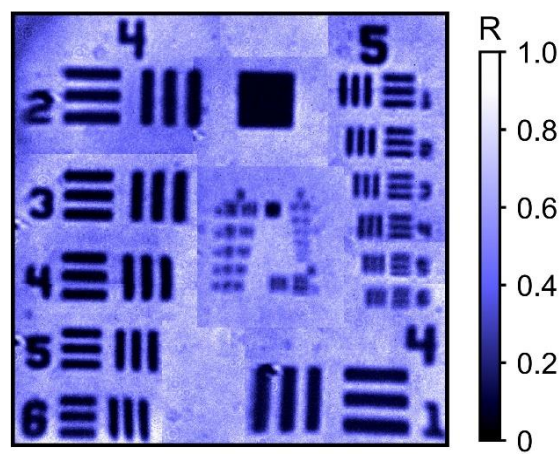

**Figure S2.** Resolution testing. With the imaging configuration described in Methods, it is possible to resolve region 5-3, which corresponds to the 12.4 μm.